# Modeling daily pan evaporation in humid climates using Gaussian Process Regression

**Sevda Shabani [1], Saeed Samadianfard [1], Mohammad Taghi Sattari [1], Shahaboddin Shamshirband [2,3], Amir Mosavi[4,5], Tibor Kmet[6], Annamária R. Várkonyi-Kóczy [4,6]**

[1] Department of Water Engineering, University of Tabriz, Tabriz, Iran; sevdashabani7364@gmail.com & s.samadian@tabrizu.ac.ir & mtsattar@gmail.com
[2] Department for Management of Science and Technology Development, Ton Duc Thang University, Ho Chi Minh City, Viet Nam
[3] Faculty of Information Technology, Ton Duc Thang University, Ho Chi Minh City, Viet Nam
[4] Department of Automation, Obuda University, Budapest, Hungary; amir.mosavi@kvk.uni-obuda.hu
[5] School of the Built Environment, Oxford Brookes University, Oxford OX3 0BP, UK
[6] Department of Mathematics and Informatics, J. Selye University, Komarno 94501, Slovakia; kmett@ujs.sk & koczya@ujs.sk

**Abstract:** Evaporation is one of the main processes in the hydrological cycle, and it is one of the most critical factors in agricultural, hydrological, and meteorological studies. Due to the interactions of multiple climatic factors, the evaporation is a complex and nonlinear phenomenon; therefore, the data-based methods can be used to have precise estimations of it. In this regard, in the present study, Gaussian Process Regression (GPR), Nearest-Neighbor (IBK), Random Forest (RF) and Support Vector Regression (SVR) were used to estimate the pan evaporation (PE) in the meteorological stations of Golestan Province, Iran. For this purpose, meteorological data including PE, temperature (T), relative humidity (RH), wind speed (W) and sunny hours (S) collected from the Gonbad-e Kavus, Gorgan and Bandar Torkman stations from 2011 through 2017. The accuracy of the studied methods was determined using the statistical indices of Root Mean Squared Error (RMSE), correlation coefficient (R) and Mean Absolute Error (MAE). Furthermore, the Taylor charts utilized for evaluating the accuracy of the mentioned models. The outcome indicates that the optimum state of Gonbad-e Kavus, Gorgan and Bandar Torkman stations, Gaussian Process Regression (GPR) with the error values of 1.521, 1.244, and 1.254, the Nearest-Neighbor (IBK) with error values of 1.991, 1.775, and 1.577, Random Forest (RF) with error values of 1.614, 1.337, and 1.316, and Support Vector Regression (SVR) with error values of 1.55, 1.262, and 1.275, respectively, have more appropriate performances in estimating

PE. It found that GPR for Gonbad-e Kavus Station with input parameters of T, W and S and GPR for Gorgan and Bandar Torkmen stations with input parameters of T, RH, W, and S had the most accurate performances and proposed for precise estimation of PE. Due to the high rate of evaporation in Iran and the lack of measurement instruments, the findings of the current study indicated that the PE values might be estimated with few easily measured meteorological parameters accurately.

## 1. Introduction

Evaporation is usually affected by the thermal energy and the vapor pressure gradient, which depends mainly on meteorological parameters. The evaporation from the soils, lakes, and water reservoirs, is one of the most critical processes in meteorological and hydrological sciences [5, 6, 15, 20]. The evaporation from the pan measures the combined weight of the temperature, relative humidity, wind speed, and sunny hours on the evaporation rate from the water surface, and this combination is effective on plant evapotranspiration. Hence, there is an excellent correlation between PE and plant evapotranspiration. Therefore, the plant evapotranspiration estimated by applying the coefficients on PE [1]. However, there is no possibility to install and maintain a meteorological device at any place, especially in impassable regions. The performance of the pan is affected by instrumental restrictions and operational issues such as human errors, instrumental errors, water turbidity, the animals and birds interfering with the water in the pan and maintenance problems [4]. Therefore, the need for models for accurate estimation of the evaporation felt more than ever before. In this regard, Keskin and Terzi [11] studied the meteorological data of the stations near the lake in western Turkey to determine the daily PE using the neural network model. The results revealed that the best structure of the model obtained with four input data including air temperature, water surface temperature, sunny hours and air pressure; however, wind speed and relative humidity have a low correlation with the evaporation rate in the study area. Guven and Kisi [9] studied the ability of linear genetic programming (LGP) in modeling the PE. They compared the results of this method with the results of Radial Basis Function Neural Network, Generalized regression neural network (GRNN), and the Stefanz-Stewart's model. Comparing the results showed that the LGP was more accurate than the other mentioned methods. Seydou and Aytac [17] studied the ability of Gene Expression Programming (GEP) in Burkina Faso coastal areas to model the evapotranspiration, and they used the combined meteorological data as inputs to the GEP model. They concluded that the GEP model has an excellent ability based on regional data. Gundalia and Dholakia [8] assessed the performance of six empirical models in predicting the daily PE. They found that the Yansen's model, based on the sunny hours, was the most appropriate method for evaluating the daily evaporation rate at the studied stations in India. Kisi and Zounemat-Kermani [13] compared two methods of neuro-fuzzy and Gene Expression Programming (GEP) to model the daily reference evapotranspiration of the Adana station in Turkey. In this study, the wind speed identified as an effective parameter in modeling. Wang et al. [21, 22] evaluated six machine learning algorithms as well as two empirical models for predicting monthly evaporation in different climate regions of China during the years of 1961-2000. They found that the multi-layer perceptron model by using the regional input data was better at the studied stations. Malik et al. [14] used the Multi-Layer Perceptron Model, Adaptive Neuro-Fuzzy Inference System, and Radial Basis Function Neural Network to predict evapotranspiration in two regions of Nagar and Ranichari (India). The results showed that the models of Adaptive Neuro-Fuzzy Inference System and multi-layer perceptron neural networks with six meteorological input data were better than other models for estimating the monthly evaporation. Recently, Feng et al. [7] examined the performance of two solar radiation-based models for estimation of daily evaporation in different regions of China. They suggested that Stewart's model can be preferred when the meteorological data of sunny hours and air temperature are available. Therefore, it is possible to estimate the evaporation through intrinsically nonlinear models. Qasem et al. [25] examined the applicability of wavelet support vector regression and wavelet artificial neural networks for predicting PE at Tabriz and Antalya stations. Obtained results indicated that artificial neural networks had better performances, and the wavelet transforms did not have significant effects in reducing the prediction errors at both studied stations.

The literature review showed that the data mining methods had a suitable application in estimating PE in different climates, but to the best of our knowledge, the application of Gaussian process regression not reported for estimating PE. So, in the present study, the ability of four data mining methods, including Gaussian process regression, support vector regression, Nearest-Neighbor, and Random Forest are studied in estimating PE rates, using different combinations of meteorological parameters. In the following, the results compared. Finally, using performance evaluation indices, the best method is obtained for estimating evaporation in the humid regions of Iran.

## 2. Materials and Methods

### *2.1. Study Area and used data*

Golestan province is located in the southeastern part of the Caspian Sea with an area of 20387 $km^2$ and covers about 1.3% of the total area of Iran (Figure 1). This province has an average annual rainfall of 450 mm in the geographical range of 36° 25' to 38° 8' north latitude and 53° 50' to 56° 18' east longitude. The geographical location and topography of Golestan province influenced by various climatic factors and different climates observed in this province. So that, the semi-arid climate is observed in the international border and the Atrak basin, moderate and semi-humid in the southern and western parts of the province, as well as cold climate in the mountainous regions.

Meteorological parameters which implemented at the current research are temperature (T), relative humidity (RH), wind speed (W) and sunny hours (S) and PE with the time period of 2011 to 2017. Table 1 represents the statistical parameters of all utilized variables. T and RH variables, due to their lower skewness values, show normal distributions. Additionally, in all studied stations, T and S variables have higher values of correlation with PE. Furthermore, RH has an inverse correlation with PE. Also, the distributions of all parameters in three studied stations illustrated in Figure 2.

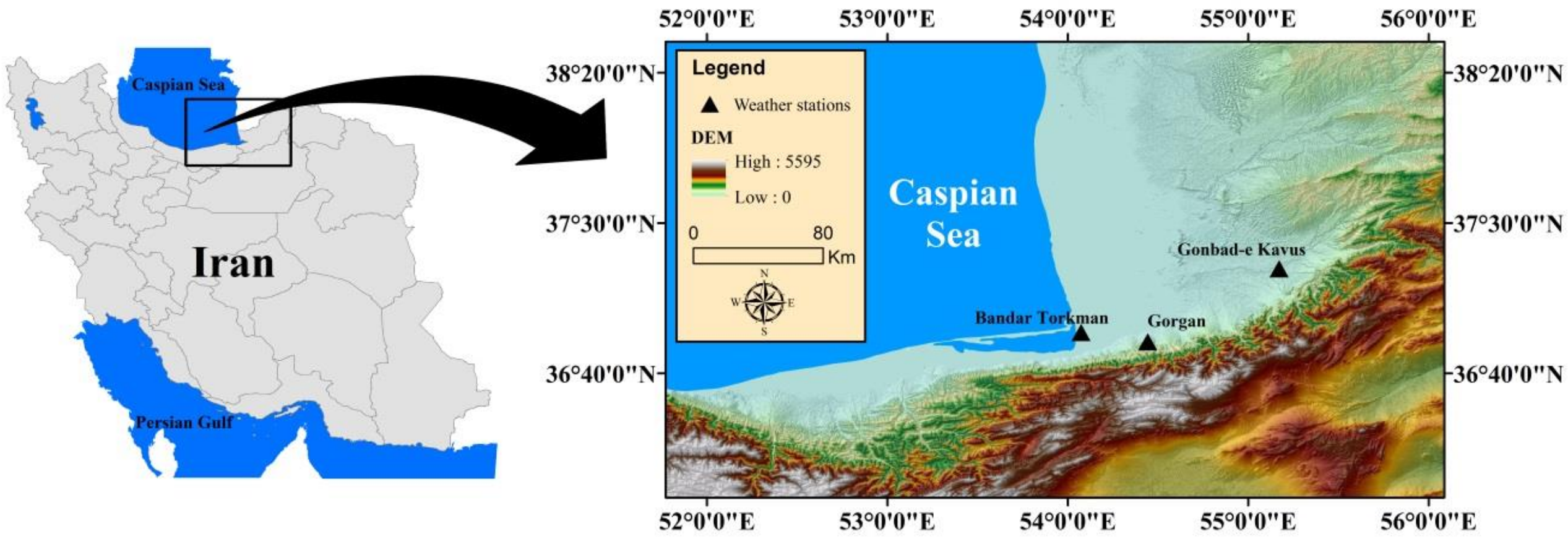

**Figure 1.** Study area

**Table 1.** Statistical characteristics of the utilized data

| Station | Variable | mean | minimum | maximum | standard deviation | coefficient of variation | skewness | Correlation with PE |
|---|---|---|---|---|---|---|---|---|
| Gonbad-e Kavus | T | 19.2 | -6.8 | 36.8 | 8.86 | 0.46 | -0.07 | 0.85 |
| | RH | 66.0 | 21.5 | 98.0 | 14.16 | 0.21 | 0.03 | -0.63 |
| | W | 1.7 | 0.0 | 7.7 | 1.07 | 0.64 | 0.72 | 0.24 |
| | S | 6.9 | 0.0 | 13.6 | 4.19 | 0.61 | -0.45 | 0.46 |
| | PE | 3.8 | 0.0 | 13.6 | 3.06 | 0.81 | 0.64 | 1.00 |
| Gorgan | T | 18.4 | -4.7 | 35.0 | 8.42 | 0.46 | -0.08 | 0.87 |
| | RH | 70.2 | 20.5 | 98.0 | 12.22 | 0.17 | -0.06 | -0.65 |
| | W | 2.0 | 0.0 | 10.0 | 1.44 | 0.71 | 0.74 | 0.28 |
| | S | 6.4 | 0.0 | 13.1 | 4.19 | 0.65 | -0.27 | 0.50 |
| | PE | 3.7 | 0.0 | 12.8 | 2.84 | 0.78 | 0.59 | 1.00 |
| Bandar Torkman | T | 18.4 | -4.3 | 34.5 | 8.08 | 0.44 | -0.11 | 0.88 |
| | RH | 73.6 | 37.5 | 98.0 | 9.73 | 0.13 | -0.12 | -0.52 |
| | W | 3.3 | 0.0 | 17.7 | 1.96 | 0.60 | 1.27 | 0.44 |
| | S | 6.5 | 0.0 | 13.3 | 4.12 | 0.63 | -0.33 | 0.45 |
| | PE | 4.4 | 0.0 | 16.0 | 3.09 | 0.71 | 0.52 | 1.00 |

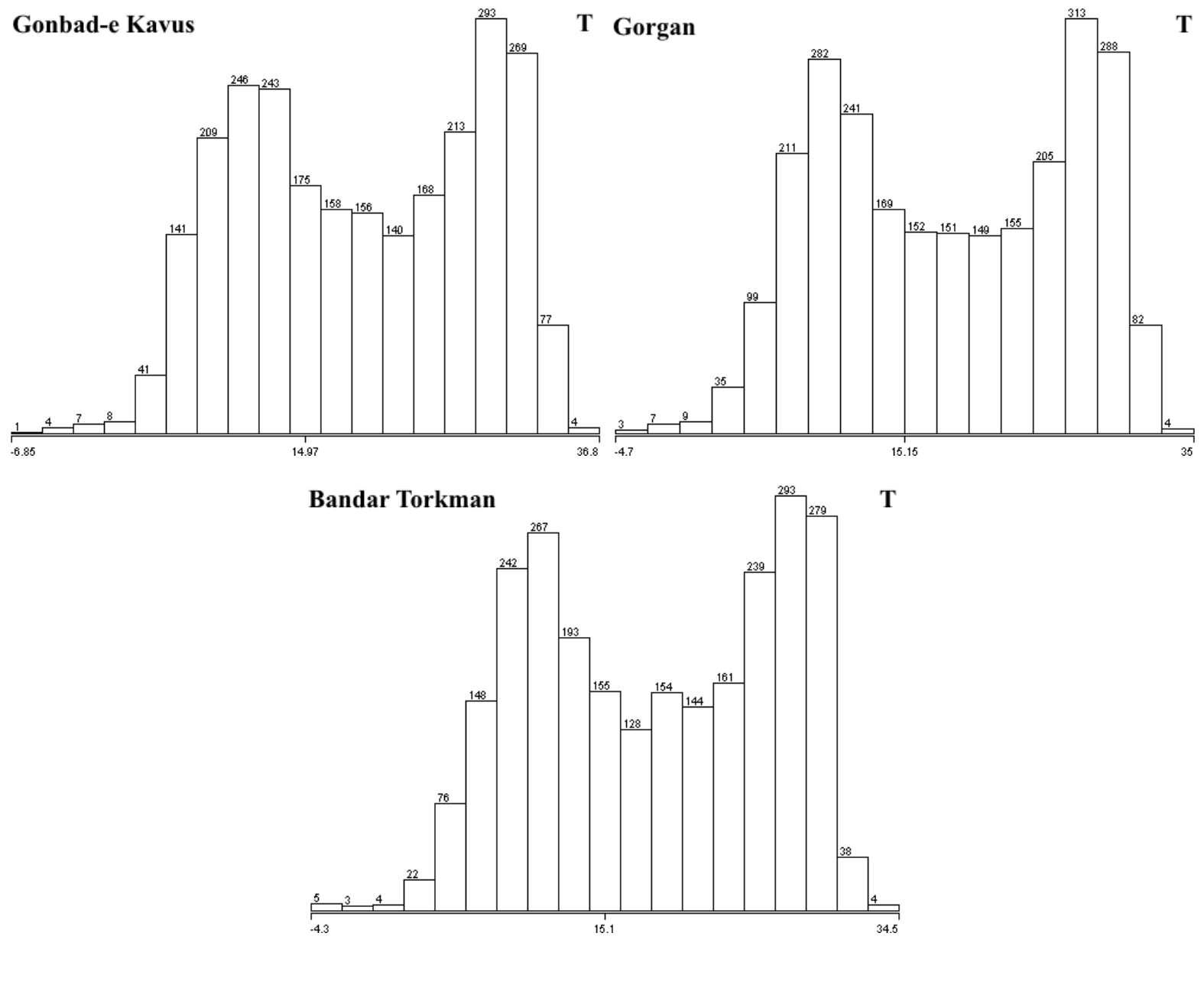

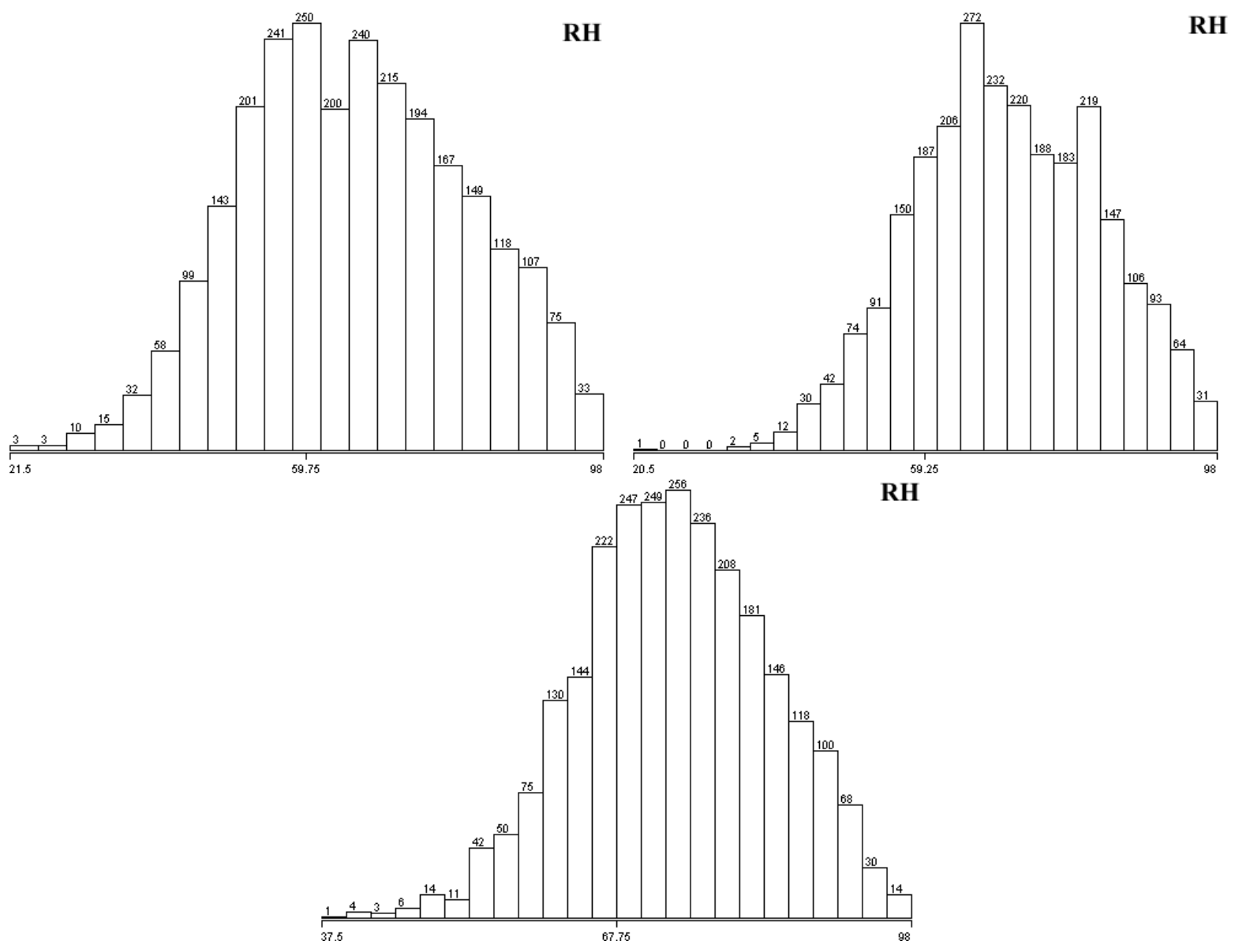

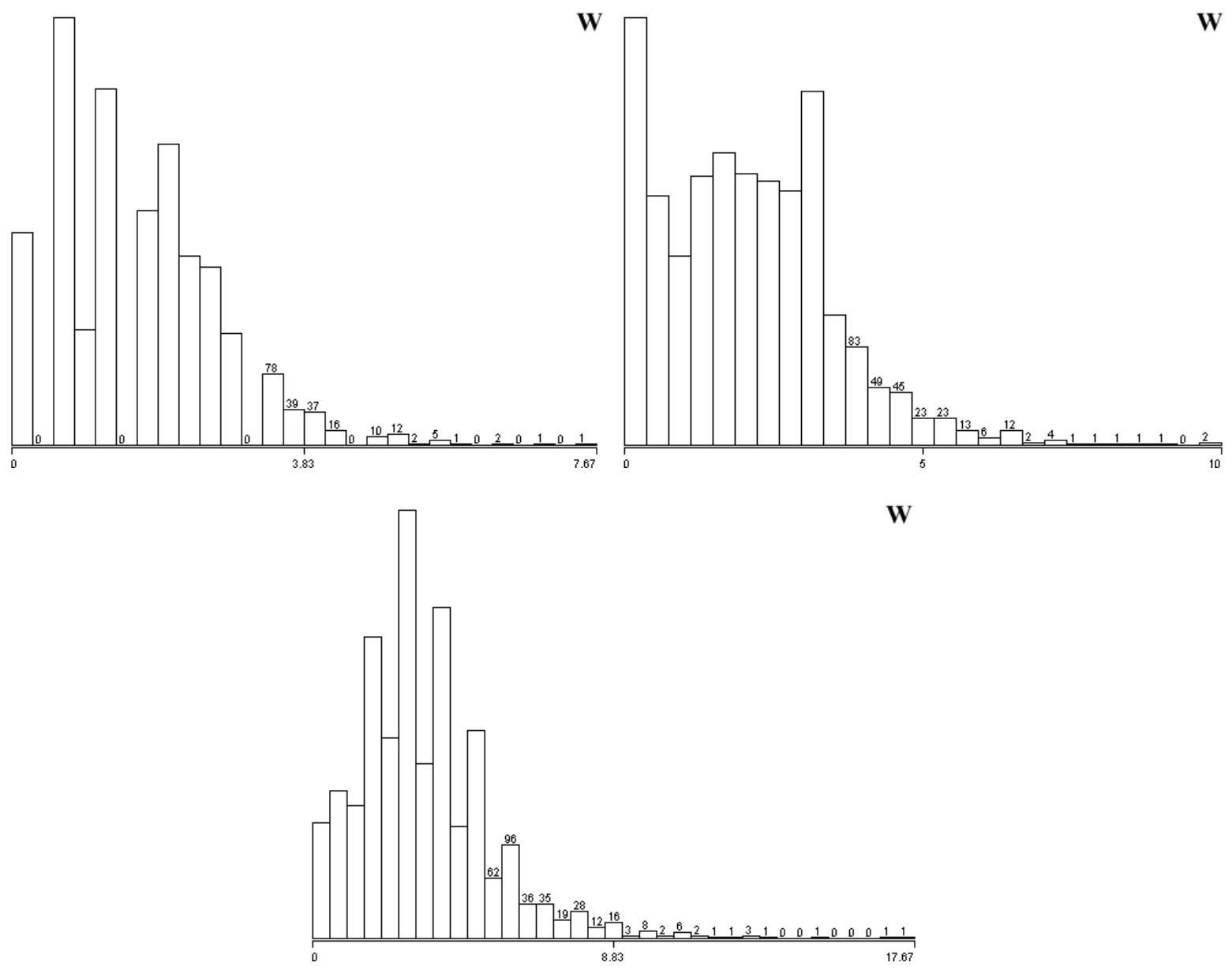

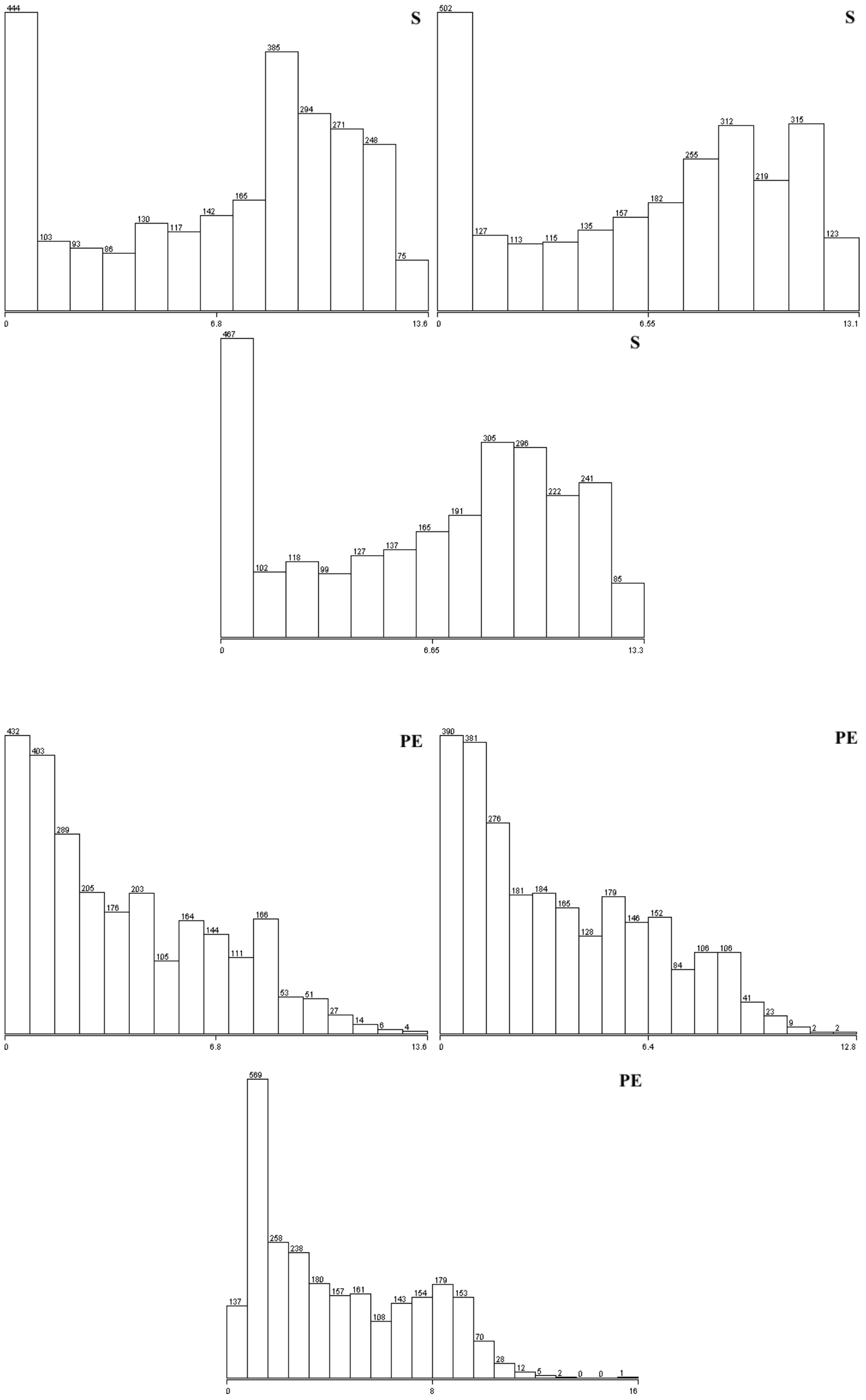

**Figure 2.** Distributions of all parameters in three studied stations

*2.2. Gaussian Process Regression (GPR)*

Gaussian processes (GP) defined as a set of random variables, in which a few variables have a multi-variable Gaussian distribution. X and Y are the input and output domains, respectively; n pairs (xi, yi) domains are independent and extracted and distributed equally. It assumed that y ∈

Re and the Gaussian process on X is defined by the average function of μ: Y → Re and the covariance function of k: X * X → Re. The primary assumption of the GPR is that y is determined by y = f (x) + ζ, in which, £ ~ N (o, σ2). In the Gaussian process regression, there is a random variable f(x) for each input variable x, that is the value of the random function f in that location. In this study, it assumed that the observation error ζ is independent and has the same distribution as the zero mean value (μ(x) = 0) and the variance (σ2), and f(x) of the Gaussian process on X (denote with k) (Eq. 1):

$$Y = (Y_1, \ldots, Y_n) \sim N(o, K = \sigma^2 I) \quad (1)$$

in which, I is the identity matrix and Kij = k(xi, xj). Because Y/X ~ N (o, K + σ2I) is normal, the conditional distribution of the test labels with the condition of the training and testing data p(Y*/Y, X, X*); in this condition (Y*/Y, X, X*) ~ N (μ, σ), therefore:

$$\mu = K(X_\circ, X)(K(X, X) + \sigma^2 I)^{-1} Y$$

(2)

$$\sigma = K(X_\circ, X_\circ) - \sigma^2 I - K(X_\circ, X)(K(X, X) + \sigma^2 I)^{-1} K(X, X_\circ) \quad (3)$$

There are some n* test data and n training data. Thus, K(x, x′) is the matrix n × n* of the evaluated covariance in all pairs of test and training data sets that are similar for other values of K(X, X), K(X, X*), and K(X*, X*). Where X and Y are the training vector and training data labels yi, while X* is the test data. The covariance function specified for creating a semi-finite positive covariance matrix of k, in which Kij = k(xi, xj). By specifying the kernel k and the noise degree σ2, eqs. 3 and 4 are sufficient for a deduction. The selection of the appropriate covariance function and its parameters is essential during the training process of GPR models because the central role in the Gaussian process regression model belongs to the covariance function K(x, x′). This function embeds the geometric structure of the training samples. Through this function, the definition of the prior knowledge of the output function of F(0) is possible [16].

### *2.3. Nearest-Neighbor (IBK)*

In general, this algorithm used for two purposes: 1. estimation of the density function of the distribution of test data and 2. classification of the test data based on the test patterns. The first step in applying this algorithm is to find a method and a relationship to calculate the distance between the test and training data. Euclidean distance is usually used to determine the distance between test and training data:

$$d(X,Y) = \sqrt{\sum_{i=1}^{n}(x_i - y_i)^2} \qquad (4)$$

where, X represents the training data with specified parameters ($x_n$ to $x_1$), and Y represents the training data with the same number of parameters ($y_n$ to $y_1$).

$$X = (x_1, x_2, \dots, x_n) \qquad (5)$$

$$Y = (y_1, y_2, \dots, y_n) \qquad (6)$$

After determining the Euclidean distance between the data, the database samples sorted in ascending order from the least distance (maximal similarity) to the maximum distance (minimum similarity). The next step in this model is to find the number of points (k) of the experiment to estimate the characteristics of the desired database. Determining the number of neighbors (k) is one of the most critical steps, and the efficiency of this method is depended to the selection of the closest (the most similar) samples from the reference database considerably. If k is assumed to be small, the results are sensitive to the unconventional single points of the model, and if k is assumed to be significant, it is possible to place some points from other classes within the desired range. Usually, the best value for k is calculated using cross-validation [23].

### *2.4. Random Forest (RF)*

The forest classification method is a set of decision trees. In the decision tree, the next -p space of the variables is divided into a hierarchy into smaller next -p subspace, so that the data in each region has as uniform as possible. This classification pattern employed by a structure decision tree. In this tree structure, the branching point to two sub-branches is called node. The first node of the tree is called the root, and the last one is the leaf [10]. In RF, each tree grows with a self-serving sample of the original data, and in order to perform the best division, the number of m variables selected randomly by variables is searched [3]. In the RF method, the data discrepancy determined in a completely different way from the usual distance functions. The similarity of data in this method is measured based on their placement in the same leaves (the final subspaces). In the FR, the similarity between i and j (s(i, j)) defined as the ratio of the times that the two given data are in the same leaf. The forest similarity matrix is random, symmetric, and positive. This matrix converts the following transformation into a non-similar matrix:

$$d(i,j) = \sqrt{1} - s(i,j) \qquad (7)$$

The formation of the classification tree is not dependent on the values of the variables; hence, the lack of similarity of the random forest applied to the types of variables [18].

*2.5. Support Vector Regression (SVR)*

Support vector machine is one of the learning methods introduced by Bousser et al. [2] on the basis of statistical learning theory. In the following years, they introduced the theory of optimum hyper plane as linear classifiers and introduced nonlinear classifiers with the help of kernel functions [19]. The models of support vector machines are divided into two main groups: a) the classifier models of support vector machine and b) a support vector regression model. A support vector machine model is used to solve the classification problems of data in different classes, and the SVR model is used to solve the prediction problems. Regression is meant to obtain a hyper plane that is fitted to the given data. The distance from any point on this hyper plane indicates the error of that particular point. The best method suggested for linear regression is the least squares method. However, for regression issues, the use of the least squares estimator in the presence of outlier data may be completely impossible; as a result, the regressor will show poor performance. Therefore, a robust estimator that would not be sensitive to small variations in the model should be developed. In fact, a penalty function is defined as follows [12]:

$$\mathbf{L^{\varepsilon}(x,y,f)} \begin{cases} \mathbf{0} & \mathbf{if} \quad |\mathbf{y}-\mathbf{f(x)}| \le \boldsymbol{\varepsilon} \\ |\mathbf{y}-\mathbf{f(x)}| - \boldsymbol{\varepsilon} & \mathbf{otherwise} \end{cases} \qquad (8)$$

The training data sets are S = {($x_1$, $y_1$), ($x_2$, $y_2$), ..., ($x_n$, $y_n$)} and the class of the function is as f(x) = {wTx + b, w∈Rm, b∈R}. If the data deviate from the value of ε, a deficiency variable must be defined according to the value of the deviation. In accordance with the penalty function, the minimization is defined as follows

$$\mathbf{Minimize} \ \frac{1}{2}||\mathbf{w}||^2 + \mathbf{C}\sum_{\mathbf{i=1}}^{\mathbf{N}}(\xi_i + \xi_i^*)$$

$$\mathbf{Subject\ to} \begin{cases} \mathbf{f(x)} - \mathbf{y_i} \le \boldsymbol{\varepsilon} + \xi_i \\ \mathbf{f(x)} - \mathbf{y_i} \le \boldsymbol{\varepsilon} + \xi_i^* \\ \xi_i, \xi_i^* \ge \mathbf{0} \end{cases} \quad \mathbf{i} = \quad \mathbf{1, 2, 3, \ldots, n}$$

**(9)**

where, $||w||^2$ is the norm of the weight vector, and $\xi_i$ and $\xi_i^*$ are auxiliary deficiency variables, and parameter C is the coefficient of equilibrium of complexity between the machine and the number of inseparable points obtained by trial and error.

### *2.6. Evaluation parameters*

Error values between computed and observed data are evaluated by Root Mean Square Error (RMSE), Mean Absolute Error (MAE) and correlation coefficient (R) defined in e.g., [26].

Additionally, Taylor diagrams were employed for inspecting the accuracy of the implemented models. It is outstanding that in the mentioned diagram, measured and some correspondent statistical parameters are presented at the same time. Moreover, different points on a polar plot are used in Taylor diagrams for investigating the differences between observed and estimated values. Also, the CC and normalized standard deviation are indicated by azimuth angle and radial distances from the base point, respectively [24].

## 3. Results and Discussion

In order to evaluate the possibility of using different combinations of meteorological data, seven different scenarios including various meteorological data were defined for more accurate estimation of PE (Table 2). Then, these input combinations were used in data mining methods to estimate evaporation at three stations of Gonbad-e Kavus, Gorgan and Bandar Torkaman. In the current study, all data mining models were separated, where 70% of the data was used to train and the 30% was used to test the models. The general results of the computations for the defined scenarios for the above-mentioned methods are presented in Tables 3.

**Table 2.** Parameters involved in defined scenarios.

| Number | Input parameters |
|---|---|
| 1 | T and RH |
| 2 | T and W |
| 3 | T and S |
| 4 | T, RH and W |
| 5 | T, RH and S |
| 6 | T, W and S |
| 7 | T, RH, W and S |

**Table 3.** General results of the computations for the defined scenarios for GPR, IBK, RF, SVR models.

| | Ghonbad | | | Gorgan | | | Bandar Torkman | | |
|---|---|---|---|---|---|---|---|---|---|
| **Model** | **R** | **MAE** | **RMSE** | **R** | **MAE** | **RMSE** | **R** | **MAE** | **RMSE** |
| **GPR1** | 0.898 | 1.173 | 1.575 | 0.890 | 1.014 | 1.307 | 0.894 | 1.023 | 1.372 |
| **GPR2** | 0.898 | 1.170 | 1.560 | 0.884 | 1.026 | 1.337 | 0.905 | 0.973 | 1.299 |
| **GPR3** | 0.894 | 1.153 | 1.550 | 0.890 | 1.001 | 1.313 | 0.900 | 1.003 | 1.346 |
| **GPR4** | 0.903 | 1.148 | 1.545 | 0.897 | 0.980 | 1.265 | 0.907 | 0.972 | 1.294 |
| **GPR5** | 0.900 | 1.161 | 1.561 | 0.894 | 0.993 | 1.289 | 0.898 | 1.002 | 1.344 |
| **GPR6** | 0.899 | 1.128 | 1.521 | 0.897 | 0.965 | 1.265 | 0.912 | 0.939 | 1.257 |
| **GPR7** | 0.904 | 1.134 | 1.530 | 0.901 | 0.958 | 1.244 | 0.912 | 0.946 | 1.254 |
| **IBK1** | 0.795 | 1.547 | 2.069 | 0.784 | 1.434 | 1.895 | 0.820 | 1.375 | 1.840 |
| **IBK2** | 0.784 | 1.593 | 2.106 | 0.772 | 1.40 | 1.898 | 0.823 | 1.340 | 1.817 |
| **IBK3** | 0.776 | 1.585 | 2.154 | 0.788 | 1.393 | 1.865 | 0.818 | 1.391 | 1.876 |
| **IBK4** | 0.810 | 1.513 | 1.991 | 0.798 | 1.4 | 1.827 | 0.833 | 1.285 | 1.737 |
| **IBK5** | 0.789 | 1.543 | 2.100 | 0.804 | 1.343 | 1.824 | 0.835 | 1.291 | 1.763 |
| **IBK6** | 0.809 | 1.507 | 1.994 | 0.808 | 1.340 | 1.775 | 0.844 | 1.289 | 1.745 |
| **IBK7** | 0.804 | 1.521 | 2.028 | 0.799 | 1.361 | 1.841 | 0.865 | 1.179 | 1.577 |
| **RF1** | 0.859 | 1.322 | 1.755 | 0.856 | 1.149 | 1.492 | 0.865 | 1.139 | 1.546 |
| **RF2** | 0.844 | 1.379 | 1.814 | 0.832 | 1.186 | 1.590 | 0.876 | 1.092 | 1.484 |
| **RF3** | 0.865 | 1.268 | 1.703 | 0.851 | 1.155 | 1.522 | 0.877 | 1.128 | 1.502 |
| **RF4** | 0.880 | 1.239 | 1.647 | 0.875 | 1.059 | 1.387 | 0.892 | 1.023 | 1.386 |
| **RF5** | 0.877 | 1.241 | 1.673 | 0.870 | 1.082 | 1.423 | 0.887 | 1.045 | 1.419 |
| **RF6** | 0.879 | 1.225 | 1.621 | 0.879 | 1.030 | 1.374 | 0.900 | 1.007 | 1.349 |
| **RF7** | 0.886 | 1.199 | 1.614 | 0.885 | 1.011 | 1.337 | 0.903 | 0.980 | 1.316 |
| **SVR1** | 0.895 | 1.207 | 1.629 | 0.888 | 1.006 | 1.317 | 0.891 | 1.018 | 1.389 |
| **SVR2** | 0.896 | 1.184 | 1.585 | 0.883 | 1.017 | 1.340 | 0.904 | 0.971 | 1.314 |
| **SVR3** | 0.892 | 1.154 | 1.574 | 0.889 | 0.984 | 1.315 | 0.9 | 1.003 | 1.358 |
| **SVR4** | 0.901 | 1.178 | 1.59 | 0.893 | 0.982 | 1.284 | 0.906 | 0.961 | 1.298 |
| **SVR5** | 0.894 | 1.186 | 1.619 | 0.894 | 0.974 | 1.287 | 0.895 | 1.009 | 1.368 |
| **SVR6** | 0.895 | 1.129 | 1.550 | 0.895 | 0.962 | 1.278 | 0.911 | 0.943 | 1.275 |
| **SVR7** | 0.898 | 1.146 | 1.572 | 0.898 | 0.958 | 1.262 | 0.909 | 0.960 | 1.278 |

The presented results in Table 3 showed that for the GPR at Gonbad-e Kavus station, GPR6 with R = 0.899, MAE = 1.128, and RMSE = 1.521 has less error than the other GPR combinations. However, the GPR7 model with R = 0.904, MAE = 1.134 and the MAE = 1.530 requires more meteorological parameters. After GPR6, it presented the most accurate estimations of PE. On the other hand, GPR3 using two parameters of T and S, with R = 0.894, MAE = 1.153, and RMSE = 1.550 have higher accuracy than the other SVR models. Also, due to fewer implemented parameters, GPR3 can be used in the case of data deficiency with an acceptable error and high reliability. Based on the results obtained at Gorgan Station, GPR7 with meteorological data of T, RH, W, and S has the lowest error with RMSE = 1.244, MAE = 0.958, and R = 0.901 and selected as the most accurate method among the GPR models. After the GPR7, GPR6 with RMSE = 1.265, MAE = 0.965, and R = 0.897 and GPR4 with RMSE = 1.265, MAE = 0.980 and R = 0.897 with higher error than the GPR7 is in the second order. Based on the results obtained in the Bandar Torkaman station, GPR7 with RMSE = 1.254, MAE = 0.946, and R = 0.912 is selected as superior GPR model. In the next rank, the GPR6 with RMSE = 1.257, MAE = 0.939, and R = 0.912 presented precise estimations.

According to the results obtained at Gonbad-e Kavus station in the nearest-neighbor method, the IBK-4 with R = 0.810, MAE = 1.513, and RMSE = 1.991 showed better performance than the other models. This scenario has a higher MAE than the IBK-6, however, because of the low RMSE and also the higher R-values, it can be described as the best nearest-neighbor model in evaporation estimation at Gonbad-e Kavus station. Also, IBK-6 with R = 0.809, MAE = 1.507, and RMSE = 1.994 had an acceptable performance. Besides, at Gorgan Station, IBK6 with RMSE = 1.775, MAE = 1.34, and R = 0.808, by having the lowest error rate, was selected as the superior IBK model. Furthermore, IBK5 and IBK6 can be used as the second and third ranks, respectively. According to the results obtained in Bandar Torkaman, IBK-7 with RMSE

= 1.577, MAE = 1.17, and R = 0.885 shows the best result among the IBK models. In the next rank, the IBK-4 with RMSE = 1.737, MAE = 1.285, and R =0.833 presented the acceptable estimations.

According to the results of the Gonbad-e Kavus station on the RF method, the RF7 with the lowest RMSE = 1.614, the lowest MAE = 1.999, and the highest R = 0.886 showed the best results among the RF models. In the next rank, the RF6 with RMSE = 1.621, MAE = 1.225, and R = 0.879 presented relatively precise estimations. Based on the results obtained at Gorgan, the RF7 with the least error (R = 0.885, MAE = 1.011, and RMSE = 1.337) had the best performance as compared to the other RF models. Also, in Bandar Torkaman station, RF7 with RMSE = 1.316, MAE = 0.980, and R = 0.903 introduced as the superior RF model due to the lowest error rate; however, as mentioned before, the RF7 model requires many meteorological data to develop an accurate relationship between PE and meteorological data. Similar to the GPR method, the RF6 with RMSE = 1.349, MAE = 1.007, and R = 0.90 presented the best results after RF7 model. Although the RF6 model has slightly higher error than RF7, it described as the optimum RF model due to the use of the low meteorological data.

According to the computations, among the kernel functions in all SVR models, the Pearson function provided the best results. So, for Gonbad-e Kavus station, SVR6 with the least error (R = 0.895, MAE = 1.129, and RMSE = 1.55) has the best performance in comparison with other SVR models. After that, SVR7 has the least error, but it is not recommended due to the use of more meteorological parameters than the other models. On the other hand, SVR3, with the use of two parameters of T and S, is more precise than the other SVR models (with R = 0.892, MAE = 1.154, and RMSE = 1.574). Moreover, at Gorgan Station, SVR7 with RMSE = 1.262, MAE = 0.958 and R = 0.898 has the lowest error rate and considered as the superior SVR model; however, as previously mentioned, the SVR7 requires more meteorological parameters to make accurate estimations of PE. Based on the results obtained in the Bandar Torkaman station, SVR6 with the lowest RMSE = 1.275 and MAE = 0.943 and the highest R = 0.911 yield the best result among the SVR models. On the other hand, the SVR2 using only two parameters of T and W can be used with acceptable reliability and error in case of data deficiency.

Figure 3 compares the variations of the simulated evaporation rates for the superior models (GPR, IBK, RF, and SVR) with the observed evaporation rates at the verification stage. Also, the distribution patterns of the methods mentioned above shown in Figure 4.

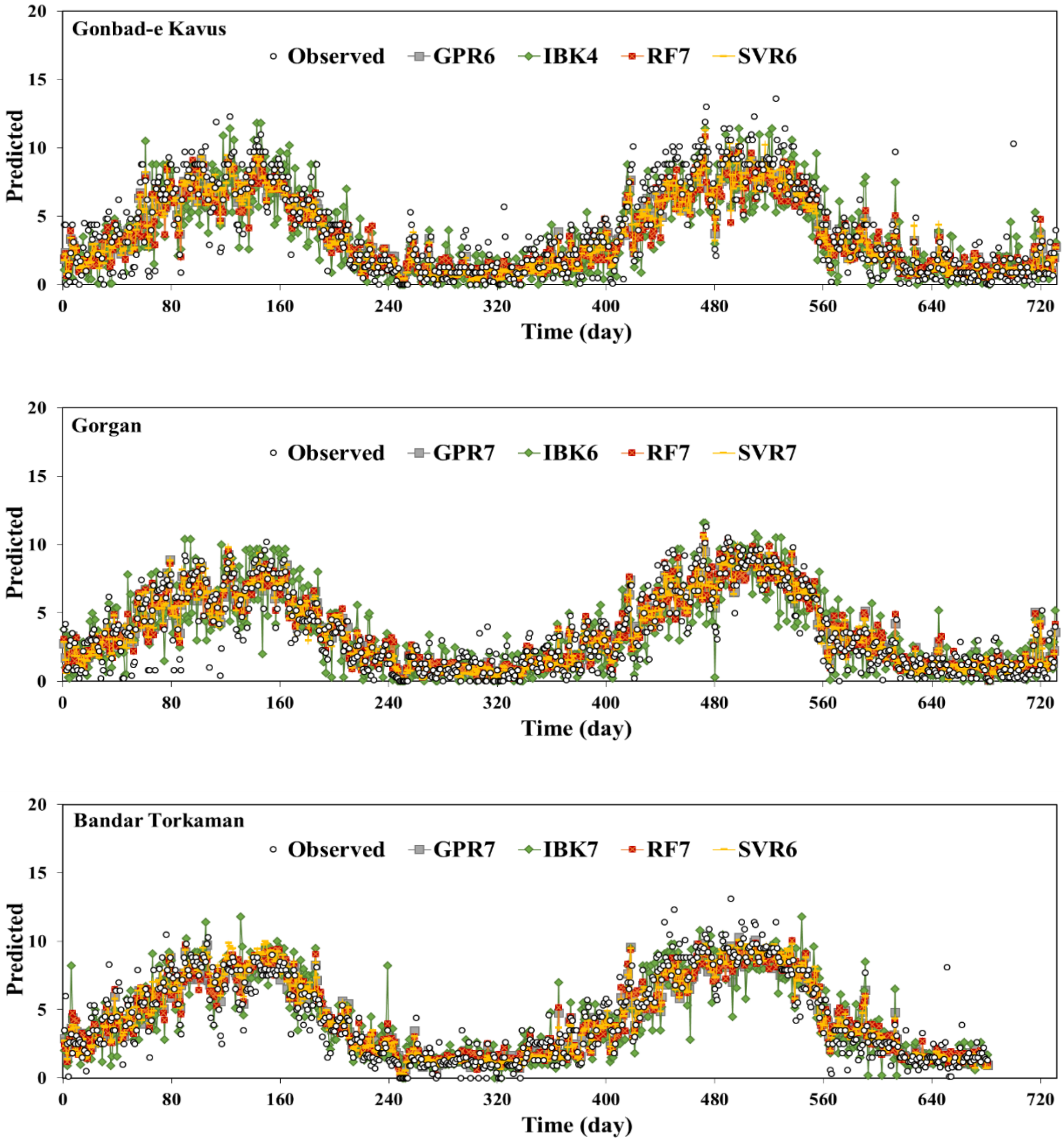

**Figure 3.** Time variations of simulated evaporation using the best models

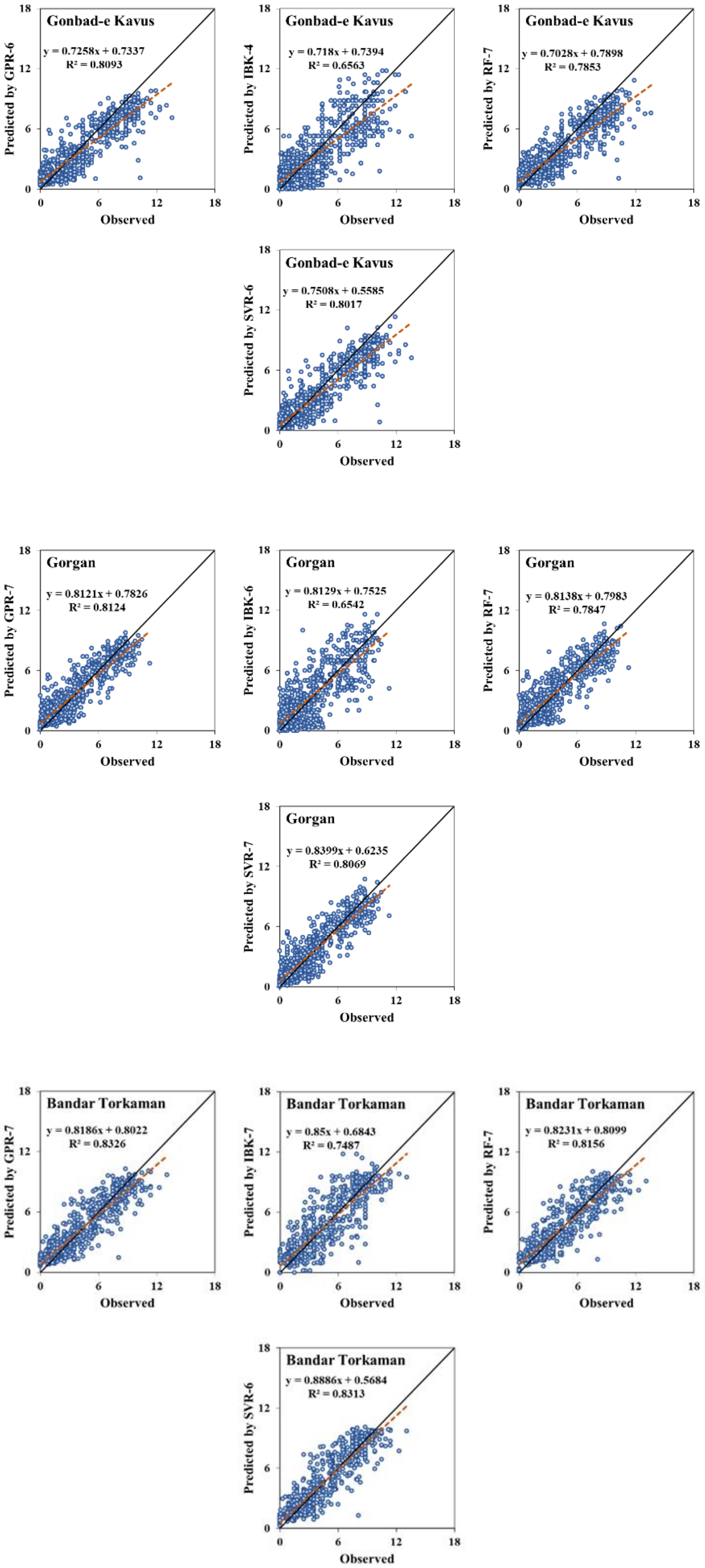

**Figure 4.** Distribution of simulated evaporation using the best models

It can be comprehended from figure 3 that the estimations of GPR6 at Gonbad-e Kavus station and GPR7 at both Gorgan and Bandar Torkman stations are in better agreement with observed PE.

Similarly, it indicates from figure 4 that the estimates of GPR6 and GPR7 are less scattered through the bisection line, and they preferred in correspondent stations. Moreover, an individual assessment of observed and estimated PE values by the best GPR, IBK, RF, and SVR models accomplished for each studied stations (Figure 5). Taylor diagrams, presented in Figure 5, are practical tools for better understanding the different potential of studied models. In the Taylor diagram, the most accurate model is explained by the point with the lower RMSE and higher R values. So, the Taylor diagrams proved that GPR6 in Gonbad-e Kavus and GPR7 in both Gorgan and Bandar Torkman stations indicated the best performances and presented the most accurate predictions of PE.

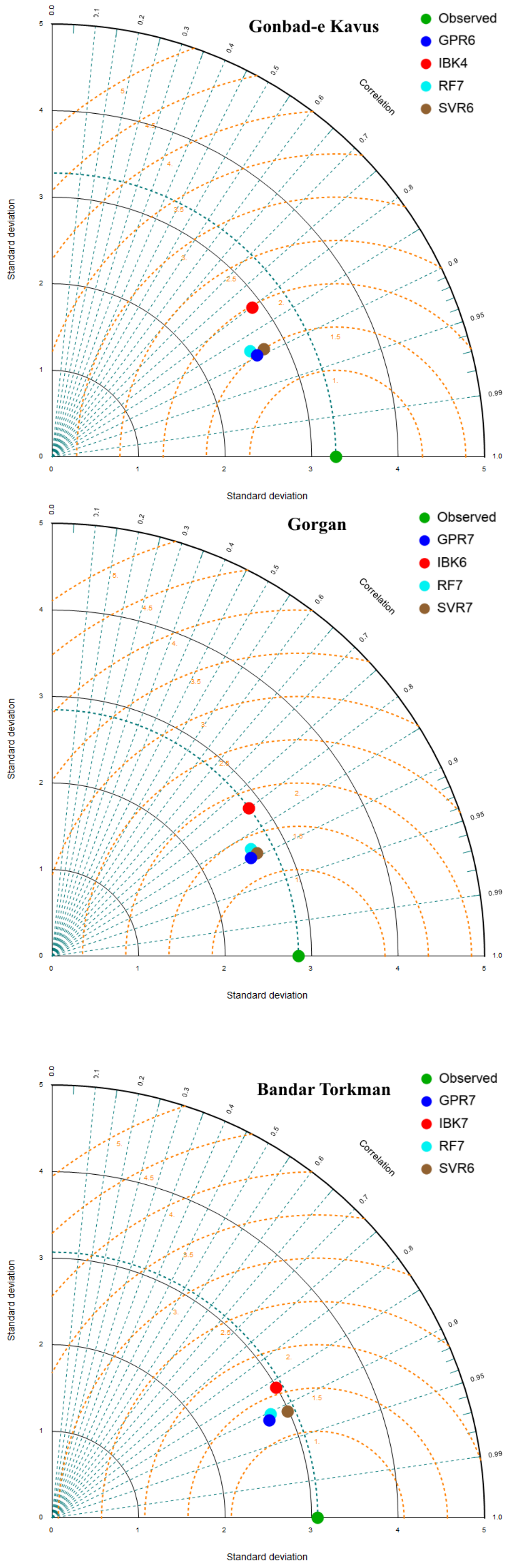

**Figure 5.** Taylor diagrams of estimated PE values

With a general look at the results and considering the above interpretations, it concludes that the meteorological parameters of T, W, and S have the most crucial role in increasing the accuracy of the PE estimation. Finally, for Gonbad-e Kavus station, GPR6 with input parameters of T, Wand S, and for Gorgan and Bandar Torkaman stations, GPR7 with input parameters of T, RH, W, and S have the best performance, and they considered as the most accurate models for estimating PE. It should be noted that the obtained results are based on the meteorological parameters of the studied stations and at a certain period, and the results changed in different climate zones.

## 5. Conclusions

Evaporation is of paramount importance in agriculture, hydrology, water and soil conservation studies. In this study, GPR, IBK, RF, and SVR methods were used to simulate daily PE in three stations of Gonbad-e Kavus, Gorgan, and Bandar Torkaman, located in Golestan province (Iran). The results of this study showed that in Gonbad-e Kavus station, the GPR6 and in the Gorgan and Bandar Torkaman stations, GPR7 have the lowest estimation errors and showed higher accuracy than other studied models. In other words, GPR can estimate PE with high accuracy using the meteorological parameters of (T, W, and S) in Gonbad-e Kavus station and the meteorological parameters of T, RH, W and S in Gorgan and Bander Turkmen stations. As a conclusive, overall result proved the superiority of the GPR method in PE estimation. The GPR recommended for PE estimation with a high degree of reliability.

## Acknowledgement

This publication has been supported by the Project: "Support of research and development activities of the J. Selye University in the field of Digital Slovakia and creative industry" of the Research & Innovation Operational Programme (ITMS code: NFP313010T504) co-funded by the European Regional Development Fund.